\documentclass[12pt,reprint,aps,groupedaddress,showpacs,nofootinbib,prd,twocolumn]{revtex4-2}

\usepackage{amsmath,amssymb,graphicx,amsfonts}
\usepackage{mathrsfs}
\usepackage{comment}
\usepackage{xcolor}
\usepackage[shortlabels]{enumitem}
\usepackage[unicode, colorlinks=true, linkcolor=red, citecolor=blue, filecolor=blue, urlcolor=blue, linktocpage, breaklinks]{hyperref}
\usepackage{url}
\usepackage{graphicx}
\usepackage{dcolumn}
\usepackage{bm}
\usepackage{wasysym}
\usepackage{color}

\begin{document}

\title{Analog Schwarzschild black holes of Bose-Einstein condensates in a cavity: \\Quasinormal modes and quasibound states}

\author{H. S. Vieira$^{1,2,3}$, Kyriakos Destounis$^{4,5}$ and Kostas D. Kokkotas$^{2}$}
\affiliation{$^1$Department of Physics, Institute of Natural Sciences, Federal University of Lavras, 37200-000 Lavras, Brazil}
\affiliation{$^2$Theoretical Astrophysics, Institute for Astronomy and Astrophysics, University of T\"{u}bingen, 72076 T\"{u}bingen, Germany}
\affiliation{$^3$S\~{a}o Carlos Institute of Physics, University of S\~{a}o Paulo, 13560-970 S\~{a}o Carlos, Brazil}
\affiliation{$^4$Dipartimento di Fisica, Sapienza Università di Roma, Piazzale Aldo Moro 5, 00185, Roma, Italy}
\affiliation{$^5$INFN, Sezione di Roma, Piazzale Aldo Moro 2, 00185, Roma, Italy} 

\begin{abstract}

Analog models of black holes have unequivocally proven to be extremely beneficial in providing critical information regarding black hole spectroscopy, superradiance, quantum phenomena and most importantly Hawking radiation and black hole evaporation; topics that have either recently begun to bloom through gravitational wave observations or have not yet been investigated in astrophysical setups. Black hole analog experiments have made astonishing steps toward the aforementioned directions and are paramount in understanding the quantum nature of the gravitational field. Recently, a tabletop analog Schwarzschild black hole has been proposed by placing Bose-Einstein condensates of photons inside a mirror's cavity, leading to a sink with a radial vortex that represents a velocity singularity. Here, we provide an extensive spectral analysis of both the tabletop acoustic black hole and its higher-dimensional gravitational analog. We find that quasinormal modes and quasibound states share qualitative similarities in both systems and show that the eikonal quasinormal modes of the analog acoustic black hole have a photon-sphere-like interpretation, which points to the existence of a phonon sphere in the analog black hole. Our results, complemented with the recently calculated graybody factors and Hawking radiation of the acoustic analog, can provide a theoretical test bed for future tabletop experiments with condensates of light in a mirror's cavity and provide significant insights regarding classical and quantum phenomena in higher-dimensional black holes.

\end{abstract}

\maketitle

\section{Introduction}\label{Intro}

The emission and detection of gravitational waves (GWs) from the binary coalescence of black holes (BHs) \cite{LIGOScientific:2016aoc,LIGOScientific:2016lio,LIGOScientific:2017ycc,LIGOScientific:2020iuh,LIGOScientific:2020stg,LIGOScientific:2020tif,LIGOScientific:2021djp} and compact objects \cite{LIGOScientific:2017vwq} have established the field of experimental gravitation and continue to provide insights into the strong field regime towards the search for an ultimate theory of gravitation. General relativity (GR) is currently been tested meticulously \cite{Berti:2015itd,Barack:2018yly,Destounis:2020kss,Destounis:2021mqv} and has proven to be the most successful to date.

GWs carry unparalleled information during all stages of the binary's evolution as it undergoes the inspiral, merger and ringdown stage. During these phases, GWs bear the externally observable properties of the binary's components and the eventual compact object's parameters that forms after ringdown, described by quasinormal modes (QNMs) \cite{Kokkotas:1999bd,Berti:2009kk,Konoplya:2011qq}. BHs undergo surprising phenomena, such as Hawking radiation when semiclassical effects are taken into account \cite{Hawking:1975vcx}, they superradiate \cite{Brito:2015oca} under the expense of the BH's spin \cite{Penrose:1971uk} or charge \cite{Bekenstein:1973mi,Cardoso:2018nvb,Destounis:2019hca,Mascher:2022pku,Destounis:2022rpk}, and when perturbed, they vibrate according to their characteristic spectra.

The BH spectroscopy program is currently aiming towards an ultimate understanding of QNMs, how to detect them properly from GW data \cite{Isi:2019aib,Giesler:2019uxc,Cotesta:2022pci}, and if nonlinearities \cite{Cheung:2022rbm,Lagos:2022otp,Mitman:2022qdl,Kehagias:2023ctr}, as well as spectral instabilities \cite{Jaramillo:2020tuu,Jaramillo:2021tmt,Destounis:2021lum,Gasperin:2021kfv,Boyanov:2022ark,Berti:2022xfj,Jaramillo:2022kuv,Konoplya:2022pbc} can interfere with QNM extraction. Nevertheless, superradiance and Hawking radiation have not been detected experimentally. To this end, different models are being used in an effort to mimic BHs and recreate such processes in controlled laboratory experiments. 

Analog gravity \cite{Visser:1997ux,Barcelo:2005fc} has the potential to provide a better understanding regarding phenomena that would otherwise elude observation in the strong field regime and when quantum effects become significant. Unruh pioneered the first theoretical BH analog \cite{Unruh:1980cg}, which can be envisioned through a fluid which on its surface, sound excitations are mapped to perturbation equations that are typically found and utilized in classical and quantum gravity investigations. The surface fluctuations endure an effective spacetime geometry that is determined by the propagation speed of these excitations and their relative speed with respect to the fluid, which enables the construction of an acoustic BH analog. Tuning the velocity of the fluid leads to the formation of an acoustic event horizon analog, beyond which sound fluctuations cannot escape the dumb hole, in analogy to infalling particles into a BH event horizon which are causally disconnected with spatial infinity.

Currently, there is an abundance of analog gravity experiments that function with fluids and superfluids \cite{Rousseaux:2007is,Weinfurtner:2010nu,Richartz:2014lda,Cardoso:2016zvz,Torres:2016iee,Patrick:2018orp,Patrick:2019kis,Torres:2019sbr,Torres:2020tzs,Euve:2021mnj}, Bose-Einstein condensates (BECs) \cite{Garay:2000jj,Steinhauer:2014dra,MunozdeNova:2018fxv,Gooding:2020scc} and optical media \cite{Drori:2018ivu}. Through these analog experiments, Hawking radiation \cite{Weinfurtner:2010nu,Steinhauer:2014dra,Euve:2015vml,MunozdeNova:2018fxv,Ornigotti:2017yqw,Euve:2021mnj}, superradiance \cite{Richartz:2012bd,Richartz:2014lda,Cardoso:2016zvz,Torres:2016iee,Ciszak:2021xlw}, QNMs and quasibound states (QBSs) \cite{Patrick:2018orp,Torres:2019sbr,Torres:2020tzs} have been experimentally observed. Furthermore, white hole analogs have been constructed \cite{Jannes:2010sa,Peloquin:2015rnl,Euve:2017gfj,Fourdrinoy:2021hmi}. These experiments are not only successful scientific achievements but also can provide unmatched insights into the future of gravitational physics.

BECs, the macroscopic ground state conglomeration  of bosonic particles at low temperatures and high densities, has been observed in a abundant variety of physical systems, though the most famous and omnipresent Bose gas; the blackbody radiation, does not show BEC phase transitions, thus novel ways have been used to construct them. This has to do with the system's photons not occupying the cavities ground state. Nevertheless, Klaers \emph{et al.} \cite{Klaers_2010} reported for the first time the observation of a BEC of photons in a two-dimensional system photon gas in a dye-filled optical microcavity \cite{Klaers_2010_white_wall}. The cavity walls provide the confining potential needed and a non-vanishing photon mass that makes it analogous to a system of two-dimensional bound massive bosons. The dye in the cavity plays the role of the stabilization factor for the photons to thermalize at the dye temperature which leads to a massively populated ground-state mode formation as the photon density increases. This is the basic principle for the formation of a BEC of photons in a cavity without the photons disappearing in the cavity walls. Further theoretical and experimental works measured the radial profile of the vortex \cite{Greveling_2018}, energy-dependent thermalization effects on these setups \cite{Marelic_2015}, that subsequently led to an extreme interest in applications of photon BECs in analog GR-related systems (see \cite{Giovanazzi_2005,Lahav:2009wx,Marino:2009}, among others).

Very recently, a novel tabletop experimental three-dimensional analog of a Schwarzschild BH was proposed with BECs of photons placed in a cavity \cite{Liao:2018avv} which leads to a sink with a two-dimensional radial vortex pointing towards the central velocity singularity, that was missing in other experiments with BECs where the analog BH horizons had a one-dimensional flow pattern that smoothly interpolates between a velocity below the speed of sound and thus did not possess a singular velocity profile. The novel proposal in \cite{Liao:2018avv} serves as an analog of a rotationally-symmetric five-dimentional Schwarzschild BH with a sonic event horizon and a velocity singularity due to the radial vortex, thus can shed light on the properties close to the region of the singularity. Since BECs of light in cavities are now routine to theoretically and experimentally construct, the aforementioned proposal from \cite{Liao:2018avv} is definitely feasible and can be constructed with the methodologies discussed above.

In Ref. \cite{Liao:2018avv}, the authors have calculated the graybody factors and Hawking radiation of the proposed experimental setup in an attempt to provide data for future experiments. In our analysis we complete their study by calculating the QNMs and QBSs for the three-dimensional BEC of light in a cavity and revisit the spectrum of five-dimensional Schwarzschild BH. We find that QNMs and QBSs share qualitative similarities in both the gravitational and the BEC analog system, thus giving more legitimacy to the analogy, and identify that the photon sphere interpretation of eikonal QNMs of the gravitational BH is shared in the analog system. Thus, the newly proposed three-dimensional acoustic BH analog possesses a phonon sphere which properties are intrinsically connected to the eikonal QNMs of phase fluctuations.

\section{Higher-dimensional Schwarzschild black holes}\label{5DSBH}

The standard four-dimensional Schwarzschild solution was generalized to higher dimensions by Tangherlini in \cite{Tangherlini:1963bw}. The higher-dimensional action, which is a generalization of the Einstein-Hilbert action, is given by
\begin{equation}
S=\frac{1}{16 \pi G_{D}}\int d^{D}x\sqrt{-g}R,
\label{eq:action_Tangherlini}
\end{equation}
where $g \equiv \det(g_{\mu\nu})$, and $G_{D}$ is the $D$-dimensional gravitational constant. In what follows, for simplicity and without loss of generality, we adopt the natural units where $G_{D} = c = \hbar = 1$. By varying the action with respect to the spacetime metric $g_{\mu\nu}$, we obtain the Einstein equations in higher dimensions
\begin{equation}
R_{\mu\nu}-\frac{1}{2}Rg_{\mu\nu}=0,
\label{eq:Einstein_1}
\end{equation}
where $R_{\mu\nu}$ and $R$ are the Ricci tensor and Ricci scalar, respectively. A static, asymptotically flat and spherically-symmetric vacuum solution of Eq.~(\ref{eq:Einstein_1}), which generalizes the Schwarzschild BH solution in $D$-dimensions, has the form
\begin{equation}
ds^{2}=-f(r)dt^{2}+f(r)^{-1} dr^{2}+r^{2} d\Omega_{D-2}^{2},
\label{eq:metric_DDSBH}
\end{equation}
where the metric function, $f(r)$, and the solid angle element, $d\Omega_{D-2}^{2}$, are given by
\begin{equation}
f(r)=1-\frac{\mathcal{M}}{r^{D-3}},
\label{eq:metric_function_DDSBH}
\end{equation}
and
\begin{align}
d\Omega_{D-2}^{2}&=d\theta_{1}^{2}+\sin^{2}\theta_{1} d\theta_{2}^{2}+\sin^{2}\theta_{1}\sin^{2}\theta_{2} d\theta_{2}^{2}+\cdots\nonumber\\
&+\left[\left(\prod_{j=1}^{D-2}\sin^{2}\theta_{j}\right)d\theta_{D-1}^{2}\right].
\label{eq:angle_DDSBH}
\end{align}
The parameter $\mathcal{M}$ is related to the BH mass $M$ through the relation
\begin{equation}
\mathcal{M}=\frac{16 \pi M}{(D-2)\Omega_{D-2}},\,\,\, \Omega_{D-2}=\frac{2\pi^{\frac{D-1}{2}}}{\Gamma(\frac{D-1}{2})},
\label{eq:mass_sphere_volume_parameter_DDSBH}
\end{equation}
where $\Omega_{D-2}$ is the volume of the $(D-2)$-dimensional unit sphere and $\Gamma(x)$ is the gamma function.

Note that the solution (\ref{eq:metric_DDSBH}) simplifies to the standard Schwarzschild metric when $D=4$. In addition, we observe that when the spacetime is higher-dimensional, the asymptotic fall-off term in the metric function $f(r)$ becomes steeper. This metric is Ricci flat and is called either the Schwarzschild-Tangherlini solution or simply the $D$-dimensional Schwarzschild BH spacetime. In this work, we will focus on the five-dimensional Schwarzschild BH, due to the explicit analogy between BECs of light in a cavity that form a three-dimensional Schwarzschild analog, as discussed in the introduction. 

The explicit line element \eqref{eq:metric_DDSBH} for $D=5$ reads
\begin{align}\nonumber
ds^{2}=&-f(r) dt^{2}+f(r)^{-1} dr^{2}\\&+r^{2}\left[d\theta^{2}
+\sin^{2}\theta\left(d\phi^{2}+\sin^{2}\phi\ d\chi^{2}\right)\right],
\label{eq:metric_5DSBH}
\end{align}
with
\begin{equation}
f(r)=1-\frac{\mathcal{M}}{r^{2}},\,\,\,\, \mathcal{M}=\frac{8M}{3\pi},\,\,\,\,\Omega_{D-2}=2\pi^{2},
\label{eq:metric_function_5DSBH}
\end{equation}
where $\theta$, $\phi$ run over the range 0 to $\pi$, and $\chi$ from 0 to $2\pi$. The causal structure of spacetime can be identified from the equation
\begin{equation}
f(r)=0=(r-r_{1})(r-r_{2}),
\label{eq:surface_equation_5DSBH}
\end{equation}
whose solutions are the event horizon, $r_{1}=\sqrt{\mathcal{M}}$, and a negative non-physical solution, $r_{2}=-\sqrt{\mathcal{M}}$.

\subsection{Scalar wave equation}\label{SWE_5DSBH}

\begin{figure*}[t]
	\includegraphics[scale=0.42]{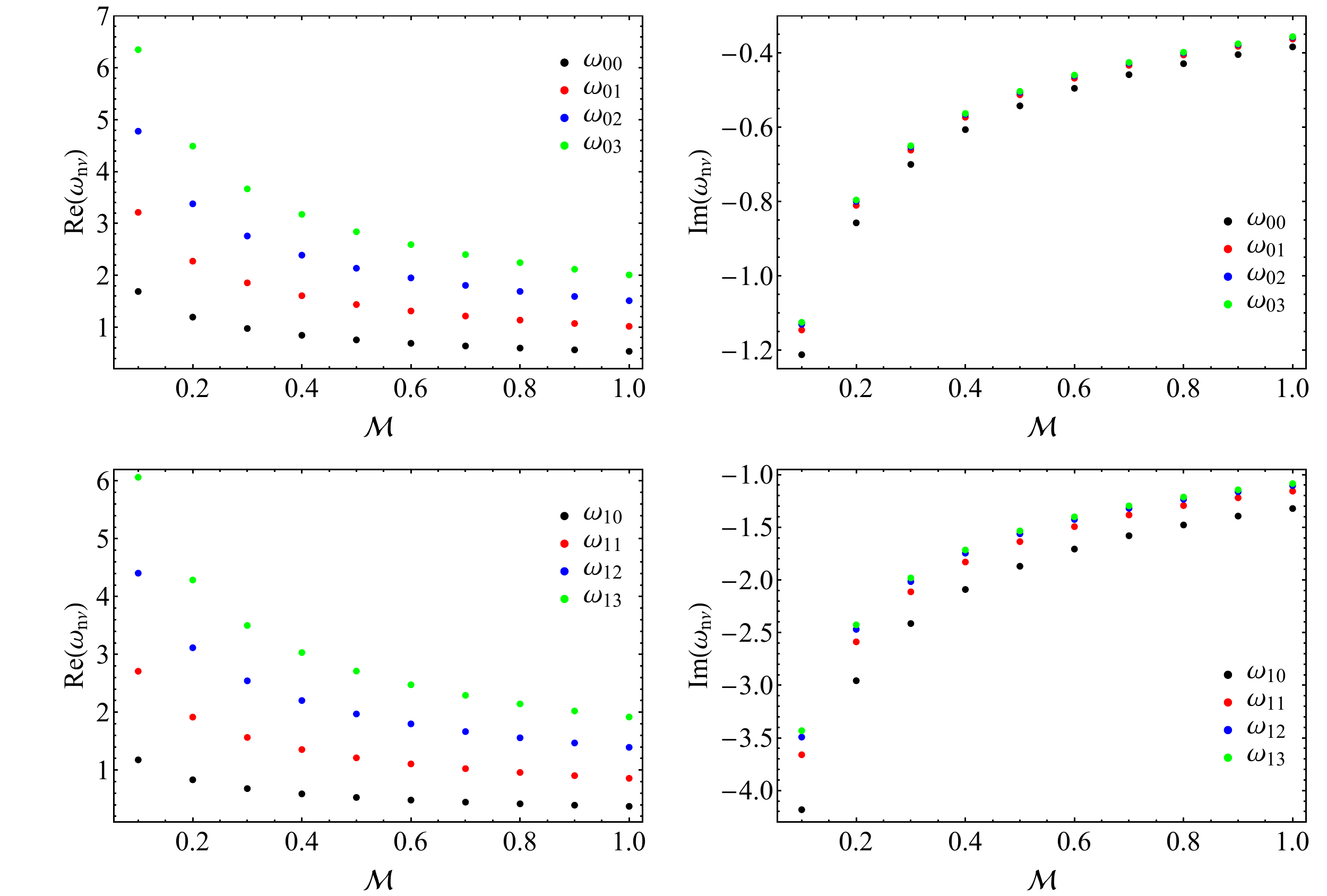}
	\caption{\emph{Top panel:} Real (left) and imaginary (right) part of the fundamental massless scalar QNMs $\omega_{0\nu}$ of a five-dimensional Schwarzschild black hole with respect to $\mathcal{M}$ and varying $\nu$. \emph{Bottom panel:} Real and imaginary part of the first overtone of massless scalar QNMs $\omega_{1\nu}$ of a five-dimensional Schwarzschild black hole with respect to $\mathcal{M}$ and varying $\nu$.}
	\label{fig:5DSBH_QNMs}
\end{figure*}

We consider a minimally-coupled massless scalar perturbation $\Psi=\Psi(t,r,\theta,\phi,\chi)$, whose equation of motion is given by the Klein-Gordon equation
\begin{equation}
\left[\frac{1}{\sqrt{-g}}\partial_{\mu}\left(g^{\mu\nu}\sqrt{-g}\partial_{\nu}\right)\right]\Psi=0.
\label{eq:Klein-Gordon_equation}
\end{equation}
Under symmetry assumptions we choose an ansatz
\begin{equation}
\Psi(t,r,\theta,\phi,\chi)=\mbox{e}^{-i \omega t}U(r)P(\theta,\phi,\chi),
\label{eq:ansatz_5DSBH}
\end{equation}
where $\omega$ is the frequency of the scalar field, $U(r)=R(r)/r^{3/2}$ is the radial function and $P(\theta,\phi,\chi)$ is an angular function. By utilizing the spacetime metric \eqref{eq:metric_5DSBH}, Eq. (\ref{eq:Klein-Gordon_equation}) can be separated into an angular and a radial equation as
\begin{align}\nonumber
&\left(\frac{1}{\sin^{2}\theta}\frac{\partial}{\partial \theta}\biggl(\sin^{2}\theta \frac{\partial}{\partial \theta}\biggr)+\lambda-\frac{1}{\sin^{2}\theta}\left[\frac{1}{\sin\phi}\frac{\partial}{\partial \phi}\biggl(\sin\phi\frac{\partial}{\partial \phi}\biggr)\right.\right.\\
&\left.\left.+\bar{\lambda}+\frac{1}{\sin^{2}\phi}\frac{\partial^{2}}{\partial \chi^{2}}\right]\right)P(\theta,\phi)=0,
\label{eq:angular_equation_5DSBH}
\end{align}
and
\begin{align}\nonumber
&R^{\prime\prime}(r)+\frac{f^\prime(r)}{f(r)}R^{\prime}(r)\\
&+\left[\frac{\omega^2}{f^2(r)}-\frac{4\lambda+3f(r)+6rf^\prime(r)}{4r^2f(r)}\right]R(r)=0.
\label{eq:radial_equation_5DSBH}
\end{align}
Here, $\lambda$ and $\bar{\lambda}$ are the separation constants to be determined, and the primes generally denote differentiation with respect to $r$.

The general exact solution of the angular equation (\ref{eq:angular_equation_5DSBH}) is given by $P(\theta,\phi,\chi)=P_{\nu,4}^{l}(\cos\theta)Y_{lm}(\phi,\chi)$, where $P_{\nu,4}^{l}(\cos\theta)$ are the associated Legendre functions in four-dimensions \cite{Torres:2013,Frye:2014,Hochstadt:1986} with arbitrary degree $\nu\in \mathbb{C}$ and order $l$ such that $\lambda=\nu(\nu+2)$, and $Y_{lm}(\phi,\chi)$ are the spherical harmonics with $l$ and $m$ being the angular and magnetic quantum numbers, respectively, such that $\bar{\lambda}=l(l+1)$.

In what follows, we will revisit the QNMs of five-dimensional Schwarzschild BHs and calculate analytically the corresponding QBSs.

\subsection{Quasinormal modes}\label{QNMs_5DSBH}

Equation \eqref{eq:radial_equation_5DSBH} can be recast into a Schr\"odinger-like equation by multiplying it with $f^2(r)$ and utilizing the tortoise coordinate $r_*$, with $dr/dr_*=f(r)$, to obtain
\begin{equation}\label{eq:QNM_equation}
	\frac{d^2 R(r)}{dr_*^2}+\left(\omega^2-V_\text{BH}\right)R(r)=0,
\end{equation}
where the effective potential $V_\text{BH}$ is
\begin{equation}\label{eq:effective_potential_BH}
	V_\text{BH}\equiv f(r)\left(\frac{\nu(\nu+2)}{r^2}+\frac{3f(r)}{4r^2}+\frac{3f^\prime(r)}{2r}\right).
\end{equation}
QNMs $\omega_{n\nu}$ are a discrete set of mode solutions of the radial equation \eqref{eq:QNM_equation} with purely ingoing (outgoing) boundary conditions at the event horizon (infinity), such that
\begin{equation}
	R(r)\sim e^{-i\omega r_*},\,\, r\rightarrow r_1,\,\,\,\,\,R(r)\sim e^{i\omega r_*},\,\, r\rightarrow \infty,
\end{equation}
where $n$ is the overtone number of the frequencies. In what follows, we present some typical scalar QNMs of a five-dimensional Schwarzschild BH. We have employed three different methods, namely a pseudospectral method \cite{Jansen:2017oag}, a matrix method \cite{Lin:2016sch,Lin:2019mmf} and the Wentzel-Kramers-Brillouin (WKB) method \cite{Iyer:1986np,Iyer:1987}. We have found convergence between the pseudospectral and matrix method QNMs up to at least six digits (convergence can reach even to 20 digits when more grid points are utilized) and confirmed the validity of eikonal QNMs with the WKB approximation finding similar precision. Our results also are in perfect agreement with those reported in \cite{Cardoso:2003vt}.

Figure \ref{fig:5DSBH_QNMs} demonstrates the typical behavior of five-dimensional Schwarzschild scalar QNMs, i.e. the increment of the BH's mass term decreases the oscillation frequency and increases the perturbation's lifetime. Furthermore, the increment of the angular number $\nu$ increases the real part of QNMs while the imaginary part is also increased (in absolute value) as expected from the eikonal limit which probes the photon sphere of the BH, where photons occupy unstable circular geodesics \cite{Cardoso:2008bp}. There, the angular frequency of null geodesics is proportional to the real part of the eikonal QNM and their instability timescale is proportional to the imaginary part of QNMs with large $\nu$. Therefore, the BH QNMs in study have a typical photon sphere interpretation \cite{Cardoso:2008bp} which we will further discuss in the upcoming sections. For a numerical presentation of some QNMs we refer the reader to Table \ref{table_5D} and Ref. \cite{Cardoso:2003vt}.

\begin{table}[h]
	\centering
	\scalebox{1.25}{
		\begin{tabular}{|c| c | c |} 
			\hline
			\multicolumn{3}{|c|}{\text{5D Scwarzschild BH}} \\
			\hline
			$\nu $ & $n=0$ & $n=1$  \\ [0.5ex] 
			\hline
			1 & 1.0160 - 0.3623 i &0.8564 - 1.1576 i\\ 
			\hline
			2  & 1.5106 - 0.3575 i &1.3927 - 1.1046 i\\  
			\hline
			10  &5.5028 - 0.3539 i  &5.4689 - 1.0639 i\\  
			\hline
	\end{tabular}}
	\caption{Massless scalar QNMs of the five-dimensional (5D) Schwarzschild BH with $\mathcal{M}=1$.}
	\label{table_5D}
\end{table}

\subsection{Quasibound states}\label{QBSs_5DSBH}

In what follows, we will analytically solve the radial equation (\ref{eq:radial_equation_5DSBH}) for a five-dimensional Schwarzschild BH. QBSs  are solutions to \eqref{eq:radial_equation_5DSBH} with purely ingoing boundary conditions at the event horizon (same as QNMs) and decaying boundary conditions at infinity. Since both QNMs and QBSs are mode solutions to the same eigenvalue problem, though with different asymptotic behavior, we will also refer to QBSs as $\omega_{n\nu}$. We will see that the fact that at infinity QBSs decay exponentially enables an analytic evaluation. To find their analytic expression, we apply the VBK approach \cite{Vieira:2016ubt,Vieira:2021xqw} in order to write the radial equation (\ref{eq:radial_equation_5DSBH}) as a Heun-type differential equation \cite{Ronveaux:1995}.

We define a new radial coordinate, $x$, and a new parameter, $x_{1}$, as
\begin{equation}
x=\frac{r-r_{2}}{-r_{2}},
\label{eq:radial_coordinate_5DSBH}
\end{equation}
and
\begin{equation}
x_{1}=\frac{r_{1}-r_{2}}{-r_{2}},
\label{eq:x1_5DSBH}
\end{equation}
such that the three original singularities $(r_{2},0,r_{1})$ are moved to the points $(0,1,x_{1})$, while maintaining a regular singularity at spatial infinity. In addition, the regular singular point at $x=x_{1}$ is always located outside the unit circle $|x_{1}| > 1$. The final step is to perform an \textit{F-homotopic transformation} $R(x) \mapsto y(x)$ given by
\begin{equation}
R(x)=x^{A_{0}}(x-1)^{A_{1}}(x-x_{1})^{A_{x_{1}}}y(x),
\label{eq:F-homotopic_5DSBH}
\end{equation}
where the coefficients $A_{0}$, $A_{1}$, and $A_{x_{1}}$ are given by
\begin{eqnarray}
A_{0}			& = & -\frac{i\omega}{r_{1}-r_{2}},\label{eq:A0_5DSBH}\\
A_{1}			& = & \frac{1}{2}-\frac{1}{r_{1}}\sqrt{r_{1}\biggl(r_{1}+\frac{\lambda}{r_{2}}\biggr)},\label{eq:A1_5DSBH}\\
A_{x_{1}}	& = & -\frac{i\omega}{r_{1}-r_{2}}.\label{eq:Ax1_5DSBH}
\end{eqnarray}
Thus, by substituting Eqs.~(\ref{eq:radial_coordinate_5DSBH})-(\ref{eq:Ax1_5DSBH}) into Eq.~(\ref{eq:radial_equation_5DSBH}), we obtain
\begin{align}\nonumber
&y^{\prime\prime}(x)+\left(\frac{1+2A_{0}}{x}+\frac{2A_{1}}{x-1}+\frac{1+2A_{x_{1}}}{x-x_{1}}\right)y^{\prime}(x)\\
&+\frac{A_{3}x+A_{4}}{x(x-1)(x-x_{1})}y(x)=0,
\label{eq:radial_2_5DSBH}
\end{align}
where the coefficients $A_{3}$, and $A_{4}$ are given by
\begin{align}\nonumber
A_{3} &= -3+A_{x_{1}}+2A_{1}(1+A_{x_{1}})+A_{0}(1+2A_{1}+2A_{x_{1}})\\
&-\frac{(x_{1}-1)(x_{1}^{2}\lambda-2\omega^{2}+2x_{1}\omega^{2})}{r_{1}^{2}x_{1}^{2}},\label{eq:A3}\\\nonumber
A_{4} &= -A_{0}-A_{x_{1}}-2A_{0}A_{x_{1}}+\frac{3x_{1}}{2}-A_{1}x_{1}-2A_{0}A_{1}x_{1}\\
&+\frac{(x_{1}-1)^{2}(x_{1}^{2}\lambda+2\omega^{2})}{r_{1}^{2}x_{1}^{2}}.\label{eq:A4}
\end{align}
The radial equation (\ref{eq:radial_2_5DSBH}) has the form of a general Heun equation \cite{Ronveaux:1995}. Therefore, the exact solution for the radial part of the Klein-Gordon equation can be written as
\begin{align}\nonumber
R_{j}(x)=&x^{\frac{1}{2}(\gamma-1)}(x-1)^{\frac{1}{2}\delta}(x-x_{1})^{\frac{1}{2}(\epsilon-1)}\\
&[C_{1,j}\ y_{1,j}(x) + C_{2,j}\ y_{2,j}(x)],
\label{eq:radial_solution_5DSBH}
\end{align}
where $C_{1,j}$ and $C_{2,j}$ are constants to be determined, and $j=\{0,1,x_{1},\infty\}$ labels the solution at each singular point, which are given as follows. The pair of linearly independent solutions at $x=0$ ($r=r_{2}$) is given by
\begin{align}
y_{1,0} =\,&\mbox{HeunG}(x_{1},q;\alpha,\beta,\gamma,\delta;x),\label{eq:y10}\\\nonumber
y_{2,0} =\,&x^{1-\gamma}\mbox{HeunG}(x_{1},(x_{1}\delta+\epsilon)(1-\gamma)+q;\\
&\alpha+1-\gamma,\beta+1-\gamma,2-\gamma,\delta;x),\label{eq:y20}
\end{align}
where $\mbox{HeunG}(x_{1},q;\alpha,\beta,\gamma,\delta;x)$ denotes a general Heun function, which is analytic in the disk $|x| < 1$, and has a Maclaurin expansion given by
\begin{equation}
\mbox{HeunG}(x_{1},q;\alpha,\beta,\gamma,\delta;x)=\sum_{n=0}^{\infty}c_{n}x^{n},
\label{eq:Maclaurin_HeunG}
\end{equation}
with
\begin{eqnarray}
-qc_{0}+x_{1} \gamma c_{1}               & = & 0 \quad (c_{0}=1),\\
P_{n}c_{n-1}-(Q_{n}+q)c_{n}+X_{n}c_{n+1} & = & 0 \quad (n \geq 1),
\label{eq:recursion_HeunG}
\end{eqnarray}
and
\begin{eqnarray}
P_{n} & = & (n-1+\alpha)(n-1+\beta),\nonumber\\
Q_{n} & = & n[(n-1+\gamma)(1+x_{1})+x_{1}\delta+\epsilon],\\\nonumber
X_{n} & = & (n+1)(n+\gamma)x_{1}.
\label{eq:P_Q_X_recursion_HeunG}
\end{eqnarray}
The pair of linearly independent solutions at $x=1$ ($r=0$) is given by
\begin{align}
y_{1,1} &= \mbox{HeunG}(1-x_{1},\alpha\beta-q;\alpha,\beta,\delta,\gamma;1-x),\label{eq:y11}\\
y_{2,1} &= (1-x)^{1-\delta}\mbox{HeunG}(1-x_{1},((1-x_{1})\gamma+\epsilon)(1-\delta)\nonumber\\
&+\alpha\beta-q;\alpha+1-\delta,\beta+1-\delta,2-\delta,\gamma;1-x).\label{eq:y21}
\end{align}
The pair of linearly independent solutions at $x=x_{1}$ ($r=r_{1}$) is given by
\begin{align}
y_{1,x_{1}} &= \mbox{HeunG}\biggl(\frac{x_{1}}{x_{1}-1},\frac{\alpha\beta x_{1}-q}{x_{1}-1};\alpha,\beta,\epsilon,\delta;\frac{x_{1}-x}{x_{1}-1}\biggl),\label{eq:y1x1}\\
y_{2,x_{1}} &= \left(\frac{x_{1}-x}{x_{1}-1}\right)^{1-\epsilon}\mbox{HeunG}\left(\frac{x_{1}}{x_{1}-1},\right.\nonumber\\
&\left.\frac{(x_{1}(\delta+\gamma)-\gamma)(1-\epsilon)}{x_{1}-1}+\frac{\alpha\beta x_{1}-q}{x_{1}-1};\alpha+1-\epsilon,\right.\nonumber\\
&\left.\beta+1-\epsilon,2-\epsilon,\delta;\frac{x_{1}-x}{x_{1}-1}\right).\label{eq:y2x1}
\end{align}
The pair of linearly independent solutions at $x=\infty$ ($r=\infty$) is given by
\begin{align}
y_{1,\infty} &= x^{-\alpha}\mbox{HeunG}\left(\frac{1}{x_{1}},\alpha(\beta-\epsilon)+\frac{\alpha}{x_{1}}(\beta-\delta)\right.\nonumber\\
&\left.-\frac{q}{x_{1}};\alpha,\alpha-\gamma+1,\alpha-\beta+1,\delta;\frac{1}{x}\right),\label{eq:y1i}\\
y_{2,\infty} &= x^{-\beta}\mbox{HeunG}\left(\frac{1}{x_{1}},\beta(\alpha-\epsilon)+\frac{\beta}{x_{1}}(\alpha-\delta)\right.\nonumber\\
&\left.-\frac{q}{x_{1}};\beta,\beta-\gamma+1,\beta-\alpha+1,\delta;\frac{1}{x}\right).\label{eq:y2i}
\end{align}
In these solutions, the parameters $\alpha$, $\beta$, $\gamma$, $\delta$, $\epsilon$, and $q$ are given by
\begin{align}
\alpha	&= \frac{1}{r_{1}-r_{2}}\left[3r_{1}-\sqrt{r_{1}\biggl(r_{1}+\frac{\lambda}{r_{2}}\biggr)}\right.\nonumber\\
&\left.+r_{2}\biggl(\sqrt{1+\frac{\lambda}{r_{1}r_{2}}}-3\biggr)-2i\omega\right],\label{eq:alpha_5DSBH}\\
\beta	&= -\frac{1}{r_{1}-r_{2}}\left[r_{1}+\sqrt{r_{1}\biggl(r_{1}+\frac{\lambda}{r_{2}}\biggr)}\right.\nonumber\\
&\left.-r_{2}\biggl(\sqrt{1+\frac{\lambda}{r_{1}r_{2}}}+1\biggr)+2i\omega\right],\label{eq:beta_5DSBH}\\
\gamma	&= 1-\frac{2i\omega}{r_{1}-r_{2}},\label{eq:gamma_5DSBH}\\
\delta	&= 1-\frac{2}{r_{1}}\sqrt{r_{1}^{2}+\frac{\lambda r_{1}}{r_{2}}},\label{eq:delta_5DSBH}\\
\epsilon & = 1-\frac{2i\omega}{r_{1}-r_{2}},\label{eq:epsilon_5DSBH}\\
q	& = \frac{r_{1}}{r_{2}}-\frac{\lambda}{r_{2}^{2}}-1+\frac{1}{r_{2}}\biggl[\sqrt{r_{1}^{2}+\frac{\lambda r_{1}}{r_{2}}}+i\omega\biggr]\nonumber\\
&-\frac{(r_{2}+2i\omega)}{r_{1}r_{2}}\sqrt{r_{1}^{2}+\frac{\lambda r_{1}}{r_{2}}}-\frac{2i(r_{1}-r_{2}-2i\omega)\omega}{(r_{1}-r_{2})^{2}}.\label{eq:q_5DSBH}
\end{align}

The first boundary condition related to QBSs (and QNMs) requires that the radial solution is purely ingoing at the event horizon, i.e. to impose $r \rightarrow r_{1}$ (or $x \rightarrow x_{1}$) on the radial solution given by Eq.~(\ref{eq:radial_solution_5DSBH}). To do this, we use Eqs.~(\ref{eq:y1x1}) and (\ref{eq:y2x1}) to get the following asymptotic behavior
\begin{equation}
\lim_{r \rightarrow r_{1}} R_{x_{1}}(r) \sim C_{1,x_{1}}\ (r-r_{1})^{\frac{1}{2}(\epsilon-1)} + C_{2,x_{1}}\ (r-r_{1})^{-\frac{1}{2}(\epsilon-1)},
\label{eq:1st_condition_5DSBH}
\end{equation}
where all the remaining constants were included in $C_{1,x_{1}}$ and $C_{2,x_{1}}$, which will be determined afterwards. Thus, from Eq.~(\ref{eq:epsilon_5DSBH}), we can write
\begin{equation}
\lim_{r \rightarrow r_{1}} R_{x_{1}}(r) \sim C_{1,x_{1}}\ \Psi_{{\rm in},x_{1}} + C_{2,x_{1}}\ \Psi_{{\rm out},x_{1}},
\label{eq:1st_wave_5DSBH}
\end{equation}
where $\Psi_{{\rm in},x_{1}}$ and $\Psi_{{\rm out},x_{1}}$ are the ingoing and outgoing scalar wave solutions, respectively, given by
\begin{eqnarray}
\Psi_{{\rm in},x_{1}}(r>r_{1})	& = & \mbox{e}^{-i \omega t}(r-r_{1})^{-\frac{i\omega}{2\kappa_{1}}},\label{eq:1st_wave_in_5DSBH}\\
\Psi_{{\rm out},x_{1}}(r>r_{1})	& = & \mbox{e}^{-i \omega t}(r-r_{1})^{+\frac{i\omega}{2\kappa_{1}}},\label{eq:1st_wave_out_5DSBH}
\end{eqnarray}
with the surface gravity of the event horizon, $\kappa_{1}$, being defined as
\begin{equation}
\kappa_{1}=\frac{1}{2} \left.\frac{df(r)}{dr}\right|_{r=r_{1}} = \frac{r_{1}-r_{2}}{2}.
\label{eq:grav_acc_5DSBH}
\end{equation}
The ingoing boundary condition is satisfied when $C_{2,x_{1}}=0$ in Eq.~(\ref{eq:1st_wave_5DSBH}) and Eq.~(\ref{eq:radial_solution_5DSBH}). Thus, we have
\begin{equation}
\lim_{r \rightarrow r_{1}} R_{x_{1}}(r) \sim C_{1,x_{1}}\ \Psi_{{\rm in},x_{1}}.
\label{eq:1st_boundary_5DSBH}
\end{equation}

\begin{figure*}[t]
	\includegraphics[scale=0.42]{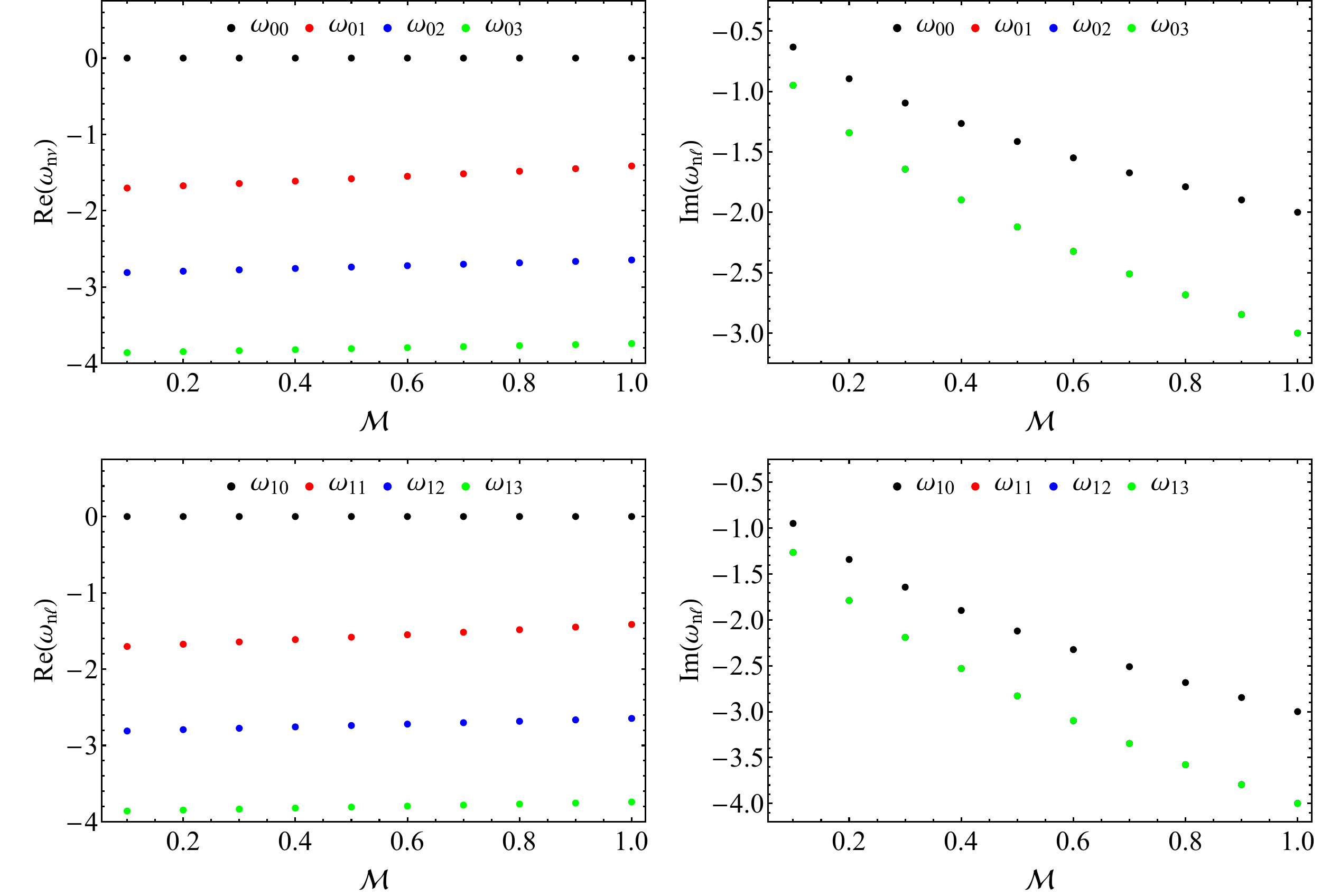}
	\caption{\emph{Top panel:} Real (left) and imaginary (right) part of the fundamental massless scalar QBSs $\omega_{0\nu}$ of a five-dimensional Schwarzschild black hole with respect to $\mathcal{M}$ and varying $\nu$. \emph{Bottom panel:} Real and imaginary part of the first overtone of massless scalar QBSs $\omega_{1\nu}$ of a five-dimensional Schwarzschild black hole with respect to $\mathcal{M}$ and varying $\nu$.}
	\label{fig:5DSBH_QBS}
\end{figure*}

The second boundary condition related to the QBSs (that differs from that of QNMs) is to require that the radial solution vanishes at asymptotic infinity, i.e. to impose a decaying boundary condition at the limit $r \rightarrow \infty$ (or $x \rightarrow \infty$) on the radial solution given by Eq.~(\ref{eq:radial_solution_5DSBH}). Thus, we use Eqs.~(\ref{eq:y1i}) and (\ref{eq:y2i}) to obtain the following asymptotic behavior
\begin{equation}
\lim_{r \rightarrow \infty} R_{\infty}(r) \sim C_{1,\infty}\ \frac{1}{r^{\sigma}},
\label{eq:2nd_condition_5DSBH}
\end{equation}
where we have imposed that $C_{2,\infty}=0$. The parameter $\sigma$ is given by
\begin{equation}
\sigma=\alpha-A_{0}-A_{1}-A_{x_{1}}.
\label{eq:sigma_5DSBH}
\end{equation}
The sign of the real part of $\sigma$ determines the asymptotic behavior of the radial scalar wave function as $r \rightarrow \infty$: if $\mbox{Re}[\sigma] > 0$, the radial solution tends to zero at spatial infinity and then it fully satisfies the second boundary condition for QBSs; if $\mbox{Re}[\sigma] < 0$, the radial solution diverges at spatial infinity.

The final asymptotic behavior of the radial scalar wave solution at spatial infinity will be determined when we know the values of the coefficients $A_{j}$, as well as the frequencies $\omega$, and the parameter $\alpha$. Those will be obtained by the VBK approach.

The general Heun functions become (class I) polynomials of degree $n$ $(\geq 0)$ if and only if they satisfy two conditions \cite{Ronveaux:1995}, namely,
\begin{eqnarray}
\alpha+n   & = & 0,\label{eq:1st_condition}\\
c_{n+1}(q) & = & 0,\label{eq:2nd_condition}
\end{eqnarray}
where the accessory parameter $q$ is given by Eq.~(\ref{eq:q_5DSBH}). The first condition, given by Eq.~(\ref{eq:1st_condition}), is called the $\alpha$-condition, which is used to find the frequency eigenvalues. The second condition, given by Eq.~(\ref{eq:2nd_condition}), determines the value of the separation constant $\lambda$ for each value of $n$, which is used to find the eigenvalues $\nu$ and both the radial and angular wave eigenfunctions. It is worth emphasizing that the two polynomial conditions could be applied in any order we wish, that is, the first condition could give the values of the separation constant, and the second condition could give the frequency eigenvalues. Thus, by imposing Eq.~(\ref{eq:1st_condition}), we obtain the exact spectrum of QBSs given by
\begin{equation}
\omega_{n\nu}=-i(\sqrt{\mathcal{M}}(n+3)-\sqrt{\mathcal{M}-\lambda}),
\label{eq:omega_5DSBH}
\end{equation}
where we have used the explicit expressions of the event horizon. Here, the separation constant is, again, $\lambda=\nu(\nu+2)$, and $n$ is the overtone number, which can be, without loss of generality, called the principal quantum number. On the other hand, by substituting Eqs.~(\ref{eq:A0_5DSBH}), (\ref{eq:A1_5DSBH}), (\ref{eq:Ax1_5DSBH}), (\ref{eq:1st_condition}) and (\ref{eq:omega_5DSBH}) into Eq.~(\ref{eq:sigma_5DSBH}), we get
\begin{equation}
\sigma=4,
\label{eq:sigma_final_5DSBH}
\end{equation}
which means that the parameter $\sigma$ is a real, positive number for any value of the principal quantum number $n$, as well as for any value of the separation constant $\lambda$, and hence the frequency eigenvalues given by Eq.~(\ref{eq:omega_5DSBH}) represent QBSs for massless scalar fields in the five-dimensional Schwarzschild spacetime. In addition, by substituting these equations into Eq.~(\ref{eq:q_5DSBH}), it is possible to obtain a nonlinear equation for $q=q(\lambda)$ and then impose the second polynomial condition given by Eq.~(\ref{eq:2nd_condition}) to obtain the angular eigenvalues $\lambda$.

Now, let us discuss a very important case of these QBSs: the fundamental mode, where $n=0$. In this case, we have
\begin{equation}
\omega_{0\nu}=-i(3\sqrt{\mathcal{M}}-\sqrt{\mathcal{M}-\lambda}),
\label{eq:omega_0nu_5DSBH}
\end{equation}
where the second polynomial condition, given by Eq.~(\ref{eq:2nd_condition}), is automatically satisfied, since it leads to
\begin{equation}
c_{n+1}(q)\biggl|_{n=0}=q=n(4+n)\biggl|_{n=0}=0.
\label{eq:q_0_5DSBH}
\end{equation}
This means that there is no restriction on the value of the separation constant $\lambda=\nu(\nu+2)$, with $\nu=0,1,2,\dots$. In particular, the ground state $(n,\nu)=(0,0)$ of QBSs is given by $\omega_{00}=-2i\sqrt{\mathcal{M}}$. This is a very important result because it designates the absence of bound states when the angular number is null. These solution are not oscillatory but only decay in time therefore no bound state can be formed in this particular case, and the BH only oscillates in accord to QNMs.

In Fig.~\ref{fig:5DSBH_QBS} we show the behavior of massless scalar QBSs in the the BH under study. The real part for the QBS $\omega_{00}$ is always zero since it does not depend on $\mathcal{M}$ and corroborates the aforementioned discussion. When $\nu\neq 0$ QBSs exist; the mass parameter acts in the opposite way with respect to that of QNMs, namely their real parts increase with $\mathcal{M}$ while their lifetime decreases. The angular number $\nu$ has a somewhat similar effect to that of QNMs, that is its increment increases (in absolute value) the oscillation frequency of QBSs and quickly drives the imaginary parts to their asymptotic, large $\nu$, value. Of course, since these modes have different boundary conditions they have no connection to null geodesics in the photon sphere at the eikonal limit.

\section{Analog black holes: Bose-Einstein condensates of light in a cavity}\label{SBHCL}

In this section, we begin by considering the general acoustic BH solution in Minkowski spacetime obtained by Unruh \cite{Unruh:1980cg} (see also Refs. \cite{Visser:1997ux,Barcelo:2005fc}), and then show how to properly choose the sound velocity in order to obtain a Schwarzschild BH in BECs of photons trapped in a mirror cavity \cite{Liao:2018avv,Wang:2019zqw}.

The fundamental equations of motion for an irrotational fluid are given by
\begin{eqnarray}
\nabla \times \mathbf{v} & = & 0, \label{eq:irrotational}\\
\partial_{t}\rho+\nabla\cdot(\rho\mathbf{v}) & = & 0, \label{eq:Continuity}\\
\rho\frac{d\mathbf{v}}{dt} \equiv \rho[\partial_{t}\mathbf{v}+(\mathbf{v}\cdot\nabla)\mathbf{v}] & = & -\nabla p, \label{eq:Euler}
\end{eqnarray}
where $\mathbf{v}$, $\rho$, and $p$ are the velocity, density, and pressure of the fluid, respectively. Next, we introduce the velocity potential $\Psi$, such that $\mathbf{v}=-\nabla\Psi$, and assume the fluid as barotropic, which means that $\rho=\rho(p)$. Then, by linearizing these equations of motion around some background $(\rho_{0},p_{0},\Psi_{0})$, namely,
\begin{eqnarray}
\rho & = & \rho_{0}+\epsilon\rho_{1}, \label{eq:rho}\\
p & = & p_{0}+\epsilon p_{1}, \label{eq:p}\\
\Psi & = & \Psi_{0}+\epsilon\Psi_{1}, \label{eq:psi}
\label{eq2:Madelung_representation}
\end{eqnarray}
we obtain the following equation
\begin{align}
&-\partial_{t}\biggl[\frac{\partial \rho}{\partial p}\rho_{0}(\partial_{t}\Psi_{1}+\mathbf{v}_{0}\cdot\nabla\Psi_{1})\biggr]+\nonumber\\
&\nabla\cdot\biggl[\rho_{0}\nabla\Psi_{1}-\frac{\partial \rho}{\partial p}\rho_{0}\mathbf{v}_{0}(\partial_{t}\Psi_{1}+\mathbf{v}_{0}\cdot\nabla\Psi_{1})\biggr]=0,
\label{eq:wave_equation_Visser}
\end{align}
where the local speed of sound, $c_{s}$, is defined by
\begin{equation}
c_{s}^{-2} \equiv \frac{\partial \rho}{\partial p}.
\label{eq:sound}
\end{equation}
Equation \eqref{eq:wave_equation_Visser} describes the propagation of the linearized scalar potential $\Psi_{1}$, that is, it governs the propagation of the phase fluctuations as weak excitations in a homogeneous stationary condensate, which can be rewritten as a wave equation in an analog curved spacetime as
\begin{equation}
\frac{1}{\sqrt{-g}}\partial_{\mu}(g^{\mu\nu}\sqrt{-g}\partial_{\nu}\Psi_{1})=0.
\label{eq:Klein-Gordon_acoustic}
\end{equation}
Note that Eq. \eqref{eq:Klein-Gordon_acoustic} is similar to the covariant Klein-Gordon equation \eqref{eq:Klein-Gordon_equation}, where the acoustic line element can be written as
\begin{equation}
ds^{2} = \frac{\rho_{0}}{c_{s}}[-c_{s}^{2} dt^{2}+(dx^{i}-v_{0}^{i} dt)\delta_{ij}(dx^{j}-v_{0}^{j} dt)].
\label{eq:acoustic_metric}
\end{equation}

Now, let us revisit how the $(1+2)$-dimensional acoustic BH was set up in Ref. \cite{Liao:2018avv}. Firstly, we can assume that the fluid is a static spherically-symmetric conservation flow, with a polar angle $\theta=\pi/2$, which implies that $d\theta=0$. Next, we define the speed of sound as $c_{s} \equiv \sqrt{kn_{0}/\mu}$, where $k$ is the strength of the effective contact interaction, $n_{0}$ is the condensate density, and $\mu$ is the effective mass of the photon gas. Then, the velocity of the fluid $\mathbf{v}$ is the flow rate along the radial direction (representing the radial vortex), that is it points to the center of the acoustic BH towards the velocity singularity, given by
\begin{equation}
\mathbf{v}_{0}=-\frac{\hbar c_{0}}{\mu r}\hat{\mathbf{r}}=-\xi c_{s}\frac{c_{0}}{r}\hat{\mathbf{r}},
\label{eq:velocity_SBHCL}
\end{equation}
where $\xi(=\hbar/\mu c_{s})$ is the correlation length, and $c_{0}(>0)$ is a constant related to the acoustic horizon's radius. Thus, the acoustic line element resulting by Eq.~(\ref{eq:acoustic_metric}) can be written as
\begin{equation}
ds^{2}=-\biggl(1-\frac{c_{0}^{2}}{r^{2}}\biggr)dt^{2}+2\frac{c_{0}}{r}dr dt+dr^{2}+r^{2}d\phi^{2},
\label{eq:line_element_SBHCL}
\end{equation}
where we have introduced the dimensionless variables $r \rightarrow \xi r$ and $t \rightarrow \xi t/c_{s}$. With the coordinate transformation
\begin{equation}
dt \rightarrow dt-\frac{c_{0}}{r(1-c_{0}^{2}/r^{2})}dr,
\label{eq:coordinate_transformation}
\end{equation}
we can write the line element of an acoustic $(1+2)$-dimensional BEC of light in a cavity as
\begin{equation}
ds^{2}=-f(r) dt^{2}+f(r)^{-1} dr^{2}+r^{2} d\phi^{2},
\label{eq:metric_SBHCL}
\end{equation}
where the warp factor $f(r)$ has the form
\begin{equation}
f(r)=1-\frac{c_{0}^{2}}{r^{2}}.
\label{eq:acoustic_metric_function}
\end{equation}
The radial vortex exists in a spatially $(1+2)$-dimensional BEC, therefore since the warp factor \eqref{eq:acoustic_metric_function} is identical to the metric function \eqref{eq:metric_function_5DSBH}, under the identification $\mathcal{M}=c_0^2$, then this model is an analog of a $(1+4)$-dimensional Schwarzschild BH. This is a consequence of the fact that the metric of the BEC does not obey Einstein's field equations, but rather it is determined by pure hydrodynamics \cite{Liao:2018avv}.

The causal structure of the particular analog BH is strikingly similar to that presented in Sec. \ref{5DSBH}. The warp factor equation
\begin{equation}
f(r)=0=(r-r_{1})(r-r_{2}),
\label{eq:surface_equation_SBHCL}
\end{equation}
has two solutions, an acoustic horizon $r_{1}=c_{0}$, which is the outermost marginally trapped surface for outgoing phonons, and an unphysical negative solution $r_{2}=-c_{0}$. The exterior acoustic event horizon is where the velocity of the fluid reaches the sound velocity. The central point of the acoustic BH can be regarded as a sink that leads to the higher-dimensional space from which the fluid flows to the three-dimensional space.

\begin{figure*}[t]
	\centering
	\includegraphics[scale=0.42]{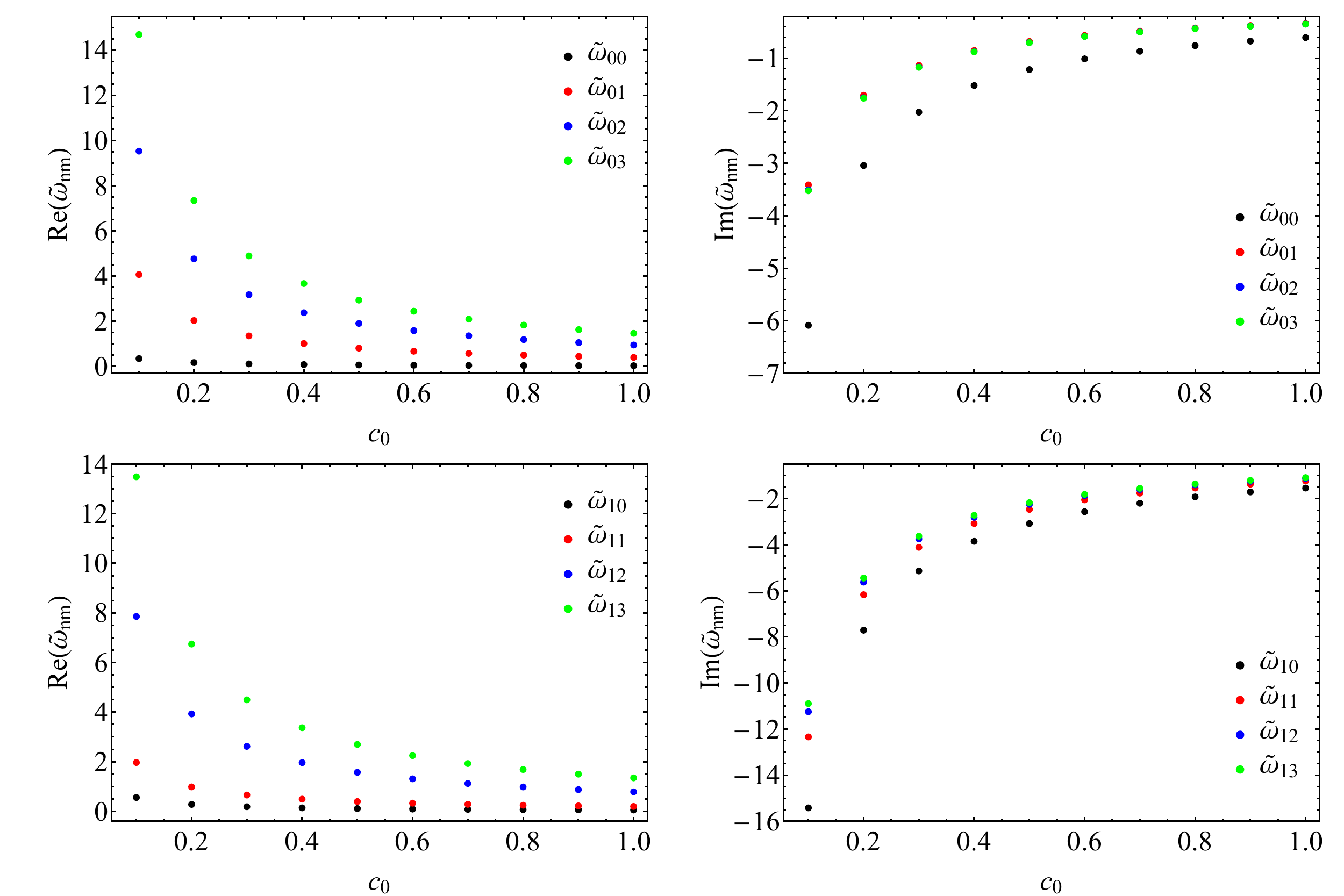}
	\caption{\emph{Top panel:} Real (left) and imaginary (right) part of the fundamental QNMs $\tilde{\omega}_{0m}$ of a three-dimensional acoustic Schwarzschild black hole with respect to $c_0$ and varying $m$. \emph{Bottom panel:} Real and imaginary part of the first overtone of QNMs $\tilde{\omega}_{1m}$ of the same system with respect to $c_0$ and varying $m$.}
	\label{fig:3DSBH_QNMs}
\end{figure*}

It is worth noting that one can also obtain this effective metric, given by Eq.~(\ref{eq:metric_SBHCL}), by using the approach developed in \cite{Ge:2019our}, which is based on the Gross-Pitaevskii theory \cite{Gross:1961,Pitaevskii:1961} that describes a nonlinear complex scalar field propagating in a curved spacetime background. In this approach, we simply require a fluid flow with velocity given by Eq.~(\ref{eq:velocity_SBHCL}) in a flat Minkowski spacetime. The proof is provided in Appendix \ref{GP}.

In what follows, we will calculate, for the first time, the QNMs and QBSs of the analog BH proposed in Ref. \cite{Liao:2018avv}, by using the same methods used for the five-dimensional Schwarzschild BH analysis in Sec. \ref{5DSBH}.

\subsection{Scalar wave equation}\label{SWE_SBHCL}

In order to solve the equation of motion given by Eq.~(\ref{eq:Klein-Gordon_acoustic}), for the acoustic metric Eq.~(\ref{eq:metric_SBHCL}), we will use a similar separation ansatz for the phase fluctuation wave function
\begin{equation}
\Psi_{1}(t,r,\phi)=\mbox{e}^{-i \tilde{\omega} t}U(r)\mbox{e}^{i m \phi},
\label{eq:ansatz_SBHCL}
\end{equation}
where $\tilde{\omega}$ is the frequency of phase fluctuations, $U(r)=R(r)/r^{1/2}$ is the radial function, and $m$ is the magnetic quantum number. Thus, by substituting Eqs.~(\ref{eq:metric_SBHCL}) and (\ref{eq:ansatz_SBHCL}) into Eq.~(\ref{eq:Klein-Gordon_acoustic}), we obtain
\begin{align}\nonumber
&R^{\prime\prime}(r)+\frac{f^\prime(r)}{f(r)}R^{\prime}(r)\\
&+\biggl[\frac{\tilde{\omega}^2}{f^2(r)}-\frac{4m^2-f(r)+2rf^\prime(r)}{4r^2f(r)}\biggr]R(r)=0.
\label{eq:radial_equation_SBHCL}
\end{align}

\subsection{Quasinormal modes}\label{QNMs_SBHCL}

Equation \eqref{eq:radial_equation_SBHCL} can be recast into a Schr\"odinger-like equation by multiplying with $f^2(r)$ and utilizing the tortoise coordinate $r_*$, with $dr/dr_*=f(r)$, to obtain
\begin{equation}\label{eq:QNM_equation_BEC}
	\frac{d^2 R(r)}{dr_*^2}+\left(\tilde{\omega}^2-V_\text{BEC}\right)R(r)=0,
\end{equation}
where the effective potential $V_\text{BEC}$ is
\begin{equation}\label{eq:effective_potential_BEC}
	V_\text{BEC}\equiv f(r)\left(\frac{m^2}{r^2}-\frac{f(r)}{4r^2}+\frac{f^\prime(r)}{2r}\right).
\end{equation}
QNMs $\tilde{\omega}_{nm}$, similarly, form a discrete set of mode solutions to the radial equation \eqref{eq:QNM_equation_BEC} with purely ingoing (outgoing) boundary conditions at the acoustic horizon (infinity), such that
\begin{equation}
	R(r)\sim e^{-i\tilde{\omega} r_*},\,\, r\rightarrow r_1,\,\,\,\,\,R(r)\sim e^{i\tilde{\omega} r_*},\,\, r\rightarrow \infty.
\end{equation}
In what follows, we present the QNMs of the three-dimensional BH analog of BECs in a cavity, where we have employed the same methods to numerically obtain and benchmark our results. 

\begin{table}[b]
	\centering
	\scalebox{1.25}{
		\begin{tabular}{||c| c | c ||} 
			\hline
			\multicolumn{3}{||c||}{\text{3D BEC analog}} \\
			\hline
			$m$ & $n=0$ & $n=1$  \\ [0.5ex] 
			\hline
			1 & 0.4068 - 0.3412 i &0.1968 - 1.2341 i\\ 
			\hline
			2  &0.9527 - 0.3507 i &0.7856 - 0.2234 i\\  
			\hline
			10  &4.9906 - 0.3535 i  &4.9532 - 1.1248 i\\  
			\hline
	\end{tabular}}
	\caption{QNMs of the three-dimensional (3D) BEC acoustic BH with $c_0=1$.}
	\label{table_3D}
\end{table}

Figure \ref{fig:3DSBH_QNMs} demonstrates the behavior of the BEC BH analog's QNMs. The increment of the vortex radius $c_0$ decreases the oscillation frequency and increases the perturbation's lifetime. The increment of $m$ increases the real part of QNMs while the imaginary part is also increased (in absolute value) as expected from the eikonal limit which probes, as we will show in the following section, the instability timescale of phase fluctuations at a critical radius where phonons are trapped in unstable curcular orbits around the acoustic BH. For some typical QNMs we refer the reader to Table \ref{table_3D}.

\subsubsection{Eikonal quasinormal mode interpretation}

We are already aware that the real and imaginary part of eikonal QNMs are intrinsically connected to the angular frequency and instability timescale of null geodesics, respectively, at the photon sphere of a vast variety of BH geometries \cite{Cardoso:2008bp,Cardoso:2017soq,Destounis:2018qnb,Liu:2019lon,Destounis:2019omd,Destounis:2020pjk,Destounis:2020yav}. Our calculations demonstrate that the same behavior occurs for the acoustic BEC BH, which means that the system, though hydrodynamic in nature, has an analog photon sphere, i.e. a phonon sphere. By calculating the QNMs for large $m$ and comparing with those of the five-dimensional BH for large $\nu$ we find that both asymptote to the same value.

Specifically, the QNMs of the acoustic BH in the eikonal limit are related to the Lyapunov exponent $\lambda_0$ of phase fluctuations at a critical radius where phonons are trapped in unstable circular orbits. The Lyapunov exponent is inversely-proportional to the instability timescale of phonon orbits there, i.e. \cite{Cardoso:2008bp}
\begin{align}
	\tilde{\omega}_\text{WKB}=m\Omega_\text{ph}-i\left(n+\frac{1}{2}\right)|\lambda_0|,\label{eq:WKB_prediction}
\end{align}
where 
\begin{equation}
	\Omega_\text{ph}=\sqrt{\frac{f(r_\text{ph})}{r^2_\text{ph}}},
\end{equation}
is the angular frequency of phonons at the phonon sphere $r=r_\text{ph}$ and
\begin{equation}
	|\lambda_0|=\sqrt{-\frac{1}{2}\frac{r^2_\text{ph}}{f(r_\text{ph})}\left(\frac{d^2}{dr^2_*}\frac{f(r)}{r^2}\right)}\Bigg|_{r=r_\text{ph}}
\end{equation}
is the Lyapunov exponent or instability timescale of phonons.

\begin{figure}[t]
	\includegraphics[scale=0.34]{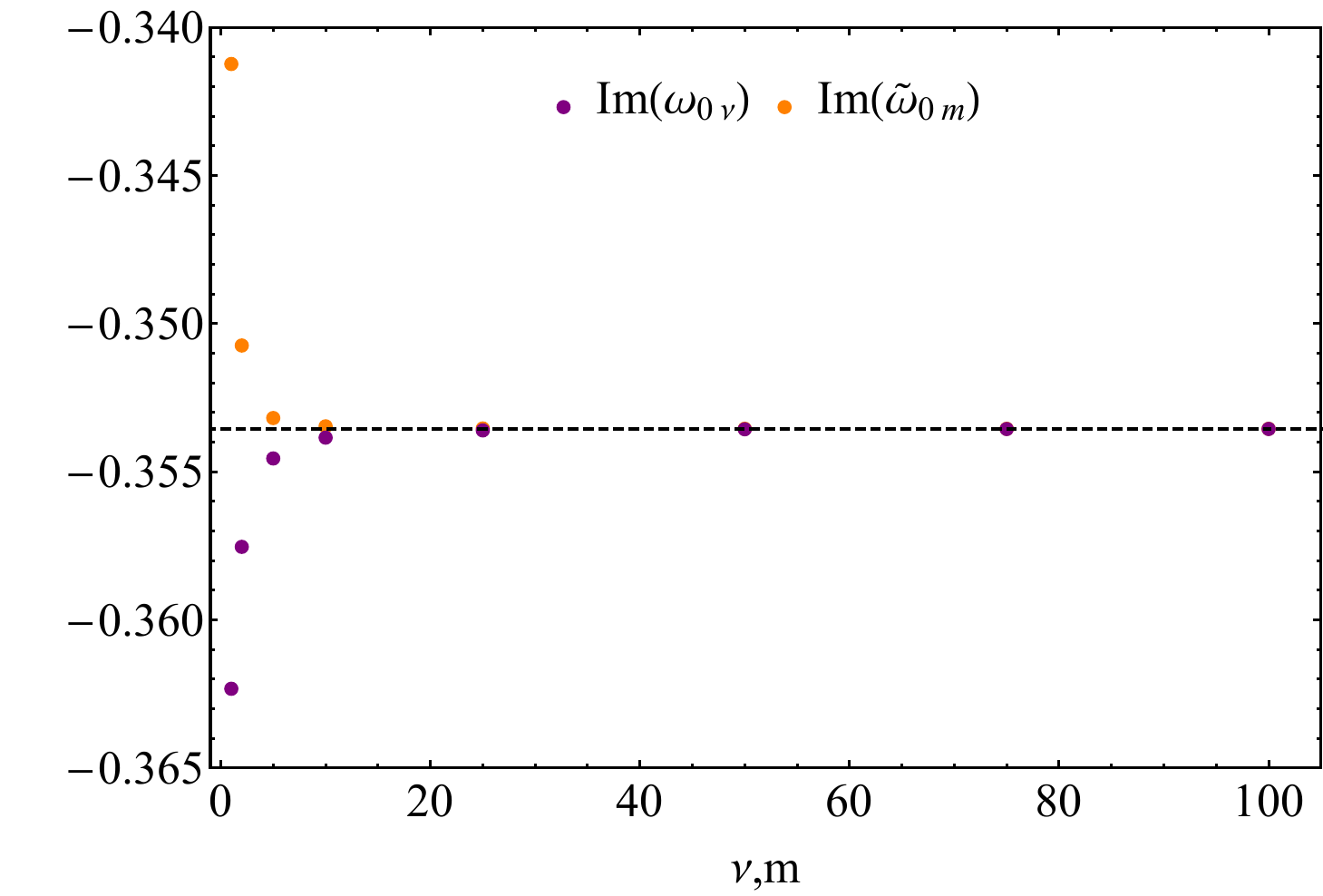}
	\caption{Imaginary parts of fundamental QNMs of a five-dimensional Schwarzschild black hole, $\text{Im}(\omega_{0\nu})$ (purple dots), with $\mathcal{M}=1$ and increasing $\nu$, and of the acoustic analog Schwarzschild black hole in a cavity, $\text{Im}(\tilde{\omega}_{0m})$ (orange dots), with $c_0=1$ and increasing $m$. The dashed horizontal line designates the WKB prediction at the eikonal limit.}
	\label{fig:eikonal_QNMs}
\end{figure}

Figure \ref{fig:eikonal_QNMs} shows that both the BH and the acoustic BEC analog reach the same decay timescale of QNMs rapidly as $\nu$ and $m$ become arbitrarily large. The asymptotic values of eikonal QNMs extracted with the numerical methods utilized above agree perfectly with those obtained by Eq. \eqref{eq:WKB_prediction} for large $m$. The fact that both the gravitational and acoustic BH asymptote to the same eikonal QNMs is routed to the form of their effective potentials. Equations \eqref{eq:effective_potential_BH} and \eqref{eq:effective_potential_BEC} are both dominated by $\nu$ and $m$ when those constants acquire large values, hence the effective potentials practically coincide and lead to the same mode solutions.

A peculiar, though not necessarily important, aspect of the convergence of the imaginary part of QNMs is that when $m$ increases the trend of convergence to the WKB prediction (shown with a dashed horizontal black line in Fig. \ref{fig:eikonal_QNMs}) is opposite to that of the BH when $\nu$ asymptotes to infinity.

\subsection{Quasibound states}\label{QBSs_SBHCL}

\begin{figure*}[t]
	\includegraphics[scale=0.42]{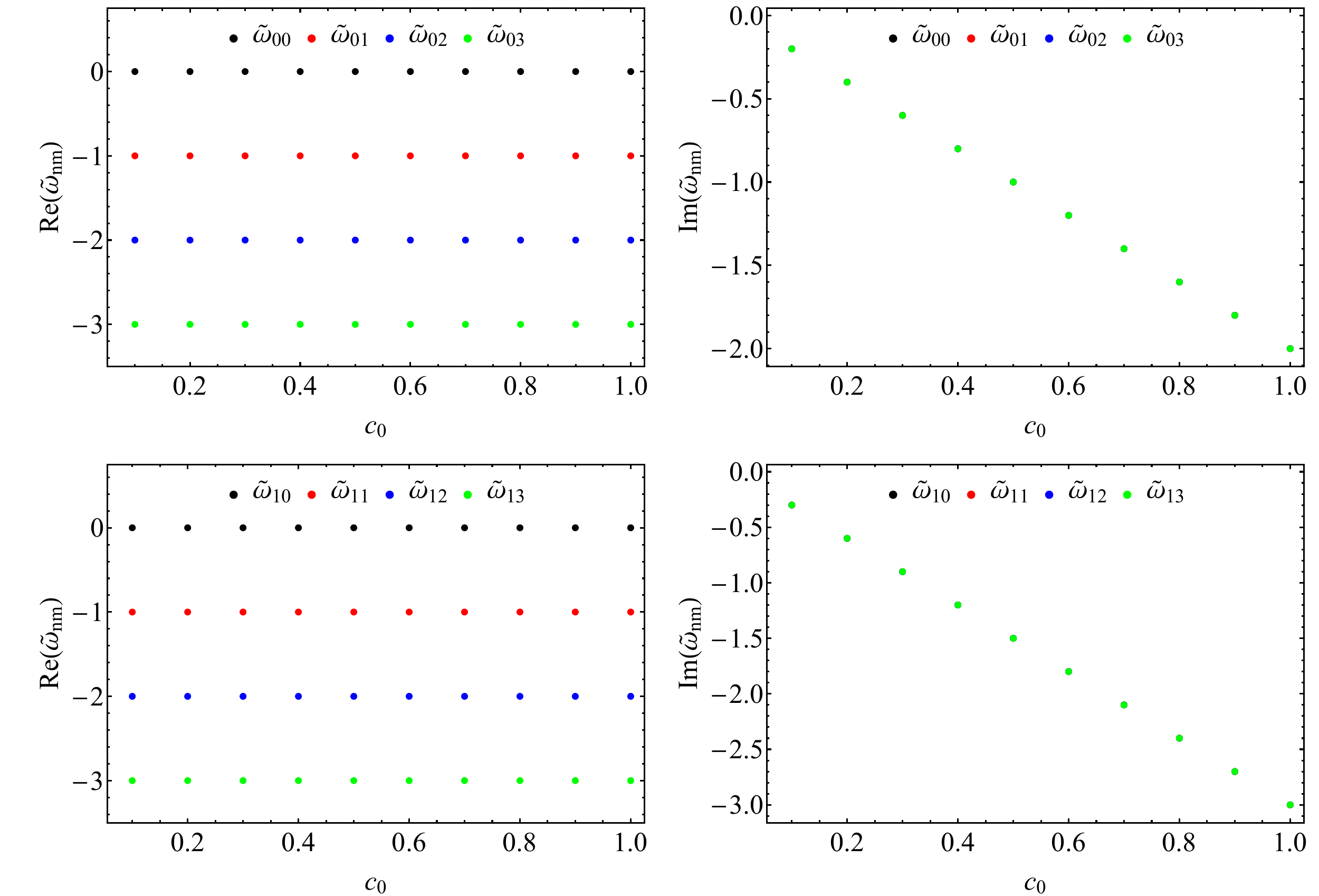}
	\caption{\emph{Top panel:} Real (left) and imaginary (right) part of the fundamental QBSs $\tilde{\omega}_{0m}$ of a three-dimensional acoustic Schwarzschild black hole with respect to the vortex radius $c_0$ and varying $m$. \emph{Bottom panel:} Real and imaginary part of the first overtone of QBSs $\tilde{\omega}_{1m}$ of the same system with respect to $c_0$ and varying $m$.}
	\label{fig:3DSBH_QBS}
\end{figure*}

The exact solution for the radial part of the massless Klein-Gordon equation, in the acoustic BH system, can be written through Eqs. (\ref{eq:radial_solution_5DSBH})-(\ref{eq:y2i}), with the new radial coordinate, $x$, the new parameter, $x_{1}$, and the \textit{F-homotopic transformation}, $R(x) \mapsto y(x)$, defined by Eqs.~(\ref{eq:radial_coordinate_5DSBH}), (\ref{eq:x1_5DSBH}), and (\ref{eq:F-homotopic_5DSBH}), respectively. However, in the present case, the coefficients $A_{0}$, $A_{1}$, $A_{x_{1}}$, $A_{3}$, and $A_{4}$ are given by
\begin{eqnarray}
A_{0}			& = & -\frac{i\tilde{\omega}}{2c_{0}},\label{eq:A0_SBHCL}\\
A_{1}			& = & \frac{1}{2}-\frac{im}{c_{0}},\label{eq:A1_SBHCL}\\
A_{x_{1}}	& = & -\frac{i\tilde{\omega}}{2c_{0}},\label{eq:Ax1_SBHCL}\\
A_{3}     & = & -\frac{(m+\tilde{\omega})(2ic_{0}+m+\tilde{\omega})}{c_{0}^{2}},\label{eq:A3_SBHCL}\\
A_{4}     & = & \frac{(m+\tilde{\omega})(2ic_{0}+m+\tilde{\omega})}{c_{0}^{2}},\label{eq:A4_SBHCL}
\end{eqnarray}
such that the parameters $\alpha$, $\beta$, $\gamma$, $\delta$, $\epsilon$, and $q$ are given by
\begin{eqnarray}
\alpha		& = & 2-\frac{i(m+\tilde{\omega})}{c_{0}},\label{eq:alpha_SBHCL}\\
\beta			& = & -\frac{i(m+\tilde{\omega})}{c_{0}},\label{eq:beta_SBHCL}\\
\gamma		& = & 1-\frac{i\tilde{\omega}}{c_{0}},\label{eq:gamma_SBHCL}\\
\delta		& = & 1-\frac{2im}{c_{0}},\label{eq:delta_SBHCL}\\
\epsilon	& = & 1-\frac{i\tilde{\omega}}{c_{0}},\label{eq:epsilon_SBHCL}\\
q					& = & -\frac{(m+\tilde{\omega})(2ic_{0}+m+\tilde{\omega})}{c_{0}^{2}}.\label{eq:q_SBHCL}
\end{eqnarray}

Similarly to Sec. \ref{QBSs_5DSBH}, by imposing ingoing boundary conditions at the acoustic horizon, decay at infinity, and the resulting $\alpha$-condition given by Eq.~(\ref{eq:1st_condition}), we obtain the exact spectrum of QBSs $\tilde{\omega}_{nm}$ in the BEC Schwarzschild analog
\begin{equation}
\tilde{\omega}_{nm}=-m-ic_{0}(2+n).
\label{eq:omega_SBHCL}
\end{equation}
Furthermore, from Eqs. (\ref{eq:A0_SBHCL})-(\ref{eq:omega_SBHCL}), we get
\begin{equation}
\sigma=\frac{i(m+ic_{0}n+\tilde{\omega}_{nm})}{c_{0}}=2,
\label{eq:sigma_final_SBHCL}
\end{equation}
which means that the parameter $\sigma$ is a real, positive (constant) number for any value of the overtone number $n$, as well as for any value of the magnetic quantum number $m$, and hence the frequency eigenvalues given by Eq.~(\ref{eq:omega_SBHCL}) represent QBSs for phase fluctuations in the acoustic BEC BH studied here.

By focusing on the fundamental mode ($n=0$) we obtain
\begin{equation}
\tilde{\omega}_{0m}=-m-2ic_{0},
\label{eq:omega_0m_SBHCL}
\end{equation}
where the second polynomial condition, given by Eq.~(\ref{eq:2nd_condition}), is automatically satisfied, since it leads to
\begin{equation}
c_{n+1}(q)\biggl|_{n=0}=q=n(2+n)\biggl|_{n=0}=0.
\label{eq:q_0_SBHCL}
\end{equation}
This means that there is no restriction on the value of the magnetic quantum number $m$; we can consider it as $m=-\infty,\dots,0,\dots,+\infty$. It is noteworthy to point that as in the BH case, the QBSs of the acoustic BH do not exist when $m=0$ since the real part is zero and the modes only decay in time, i.e. $\tilde{\omega}_{00}=-2ic_{0}$.

Fig. \ref{fig:3DSBH_QBS} demonstrates the behavior of QBSs in the analog BEC BH. Surprisingly, they have almost identical tendencies as those found for five-dimensional Schwarzschild BHs which further supports the analogy between them. Most surprisingly, the imaginary parts depend linearly on the vortex radius $c_0$, which is a much more simplified version of what occurs in the BH case.

\section{Conclusions}\label{Conclusions}

In this study, we have calculated the QNMs and QBSs of five-dimensional Schwarzschild BHs which are analog models of BECs of light in a mirror cavity; an experiment recently proposed in \cite{Liao:2018avv}. We found that the spectra for both systems share striking qualitative similarities with respect to their tuning parameters, i.e. the BH mass and the cavity's radius. Most importantly, our investigation provides proof of a phonon-sphere interpretation of the acoustic BEC BH since the imaginary part of the eikonal QNMs asymptotes to the instability timescale of phonons at the phonon sphere in accord to the WKB prediction. Our results are timely, and together with the recent investigation of the graybody factors and Hawking radiation of the analog proposed in \cite{Liao:2018avv}, can prescribe numerical data for actual experimental devices that may be constructed based on such proposal. 

In fact, although a multitude of BH analog experiments have taken place and made significant breakthroughs so far, the proposal in \cite{Liao:2018avv} is quite simple and elegant, includes a velocity singularity at the center of the vortex (aspect that other BEC-based experiments lack), has tabletop dimensions and can provide intuition regarding classical and quantum aspects of the singularity, as well as a better understanding of higher-dimensional BH geometries and their properties.

An interesting extension of the analog \cite{Liao:2018avv} can be the manipulation of the nonlinear coupling of the BECs of light in order to obtain a massive Klein-Gordon-like equation for phase fluctuations through the underlying hydrodynamic equations of motion \cite{Ciszak:2021xlw} and study classical and quantum phenomena. Most importantly, a generalization of the model in \cite{Liao:2018avv} onto a rotating BEC acoustic BH analog should be of utmost importance in order to simulate phenomena that astrophysical BHs experience. We leave these directions of research for the future.

\begin{acknowledgments}
H.S.V. is partially supported by the Alexander von Humboldt-Stiftung/Foundation (Grant No. 1209836). This study was financed in part by the Conselho Nacional de Desenvolvimento Cient\'{i}fico e Tecnol\'{o}gico -- Brasil (CNPq) -- Research Project No. 150410/2022-0. It is a great pleasure to thank the Theoretical Astrophysics at T\"{u}bingen (TAT Group) for its hospitality and technical support.
K.D. acknowledges financial support provided under the European Union's H2020 ERC, Starting Grant agreement no.~DarkGRA--757480 and the MIUR PRIN and FARE programmes (GW-NEXT, CUP: B84I20000100001).
\end{acknowledgments}

\appendix

\section{Derivation of the BEC effective metric through the Gross-Pitaevskii theory}\label{GP}
In this Appendix, we obtain an effective metric describing the acoustic $(1+2)$-dimensional BEC of light in a cavity by using the the Gross-Pitaevskii (GP) theory \cite{Gross:1961,Pitaevskii:1961}.

The equation of motion of a nonlinear (complex) scalar field $\varphi$ can be obtained from the GP theory, whose action is given by
\begin{equation}
	\mathcal{S}_{\rm GP}=\int d^{4}x\ \sqrt{-g}\ \biggl(|\partial_{\mu}\varphi|^{2}+m^{2}|\varphi|^{2}-\frac{b}{2}|\varphi|^{4}\biggr),
	\label{eq:action_GP}
\end{equation}
where $b$ is a coupling constant and $m$ is a parameter which depends on the (Hawking-Unruh) temperature $T$. We assume the temperature dependence as $m^{2} \sim T-T_{c}$, where $T_{c}$ is the critical temperature of the theory, which describes the phase transitions, where $\varphi$ is the corresponding order parameter. In the limit case when $T=T_c$, we have that $m^2=0$.

Then, from the action given by Eq.~(\ref{eq:action_GP}), we get the following equation of motion:
\begin{equation}
	\Box\varphi+m^{2}\varphi-b|\varphi|^{2}\varphi=0.
	\label{eq:equation_motion_GP}
\end{equation}
An acoustic BH solution of the GP theory can be obtained by considering perturbations of the (complex) scalar field around the spacetime background, whose metric can be fixed as
\begin{equation}
	ds^{2}=g_{tt}\ dt^{2}+g_{rr}\ dr^{2}+g_{\vartheta\vartheta}\ d\vartheta^{2}+g_{\phi\phi}\ d\phi^{2}.
	\label{eq:background_LTBH_metric}
\end{equation}
Now, by using the Madelung representation, the (complex) scalar field can be written as
\begin{equation}
	\varphi=\sqrt{\rho(\mathbf{x},t)}\mbox{e}^{i\theta(\mathbf{x},t)}.
	\label{eq:Madelung_representation_app}
\end{equation}
Here, the fluid density, $\rho$, and the phase, $\theta$, are related to the background solution in the fixed spacetime, $(\rho_{0},\theta_{0})$, and to the fluctuations, $(\rho_{1},\theta_{1})$, by
\begin{eqnarray}
	\rho & = & \rho_{0}+\epsilon\rho_{1},\\
	\theta & = & \theta_{0}+\epsilon\theta_{1}.
	\label{eq:Madelung_representation_app1}
\end{eqnarray}
Thus, the leading order equation for the background fluid density reads
\begin{equation}
	b\rho_{0}=m^{2}-g^{\mu\nu}v_{\mu}v_{\nu},
	\label{eq:leading_order}
\end{equation}
where we defined the background fluid four-velocity as $v_{\mu}=(-\partial_{t}\theta_{0},\partial_{i}\theta_{0})$, with $i=r,\vartheta,\phi$. At the sub-leading order, we obtain a relativistic wave equation which governs the propagation of the phase fluctuations as weak excitations in a homogeneous stationary condensate. It is given by
\begin{equation}
	\frac{1}{\sqrt{\mathcal{-G}}}\partial_{\mu}(\sqrt{\mathcal{-G}}\mathcal{G}^{\mu\nu}\partial_{\nu}\theta_{1})=0,
	\label{eq:phase_fluctuations}
\end{equation}
where $\mathcal{G}=\det(\mathcal{G}_{\mu\nu})$, and $\theta_{1}=\theta_{1}(t,r,\vartheta,\phi)$. Note that Eq.~(\ref{eq:phase_fluctuations}) is similar to the massless Klein-Gordon equation, from which we can extract the following effective metric:
\\
\begin{small}
	$\mathcal{G}_{\mu\nu}=\frac{c_s}{\sqrt{c^2_s-v_{\mu}v^{\mu}}}
	\begin{pmatrix}
		g_{tt}(c^2_s- v_i v^i) & \vdots & -v_{t}v_{i}\cr
		\cdots\cdots\cdots\cdots & \cdot &\cdots\cdots\cdots\cdots\cdots\cdots\cr
		-v_{i}v_{t} & \vdots & g_{ii}(c^2_s-v_\mu v^\mu)\delta^{ij}+v_i v_j\cr
	\end{pmatrix}\!,$
	\label{eq:effective_metric}
\end{small}
\\
where $c_{s}$ is the speed of sound, and defined as
\begin{equation}
	c_{s}^{2} \equiv \frac{b\rho_{0}}{2}.
	\label{speed_sound}
\end{equation}
Then, we assume that $v_{t} \neq 0$, $v_{r} \neq 0$, $v_{\vartheta} = 0$, $v_{\phi} = 0$, and $g_{rr}g_{tt}=-1$, and with the coordinate transformation
\begin{equation}
	dt \rightarrow dt+\frac{v_{r}v_{t}}{g_{tt}(c_{s}^{2}-v_{i}v^{i})}dr,
	\label{eq:coordinate_transformation_app}
\end{equation}
we can write the line element corresponding to the effective metric $\mathcal{G}_{\mu\nu}$ as
\begin{eqnarray}
	ds^{2} & = & c_{s}\sqrt{c_{s}^{2}-v_{\mu}v^{\mu}}\biggl(\frac{c_{s}^{2}-v_{r}v^{r}}{c_{s}^{2}-v_{\mu}v^{\mu}}g_{tt}\ dt^{2}\nonumber\\
	&& +\frac{c_{s}^{2}}{c_{s}^{2}-v_{r}v^{r}}g_{rr}\ dr^{2}+g_{\vartheta\vartheta}\ d\vartheta^{2}+g_{\phi\phi}\ d\phi^{2}\biggr).\nonumber\\
	\label{eq:acoustic_metric_app}
\end{eqnarray}
Finally, we can completely characterize the acoustic BH spacetime, given by Eq.~(\ref{eq:acoustic_metric_app}), by choosing a spacetime background, and the components of the fluid four-velocity as well. In this work, we focus on the flat Minkowski spacetime background, whose metric is given by
\begin{equation}
	ds^{2} = -dt^{2}+dr^{2}+r^{2}\ d\vartheta^{2}+r^{2}\sin^{2}\vartheta\ d\phi^{2}.
	\label{eq:Minkowski spacetime_app}
\end{equation}
For the fluid four-velocity, we first rescale $m^{2} \rightarrow m^{2}/2c_{s}^{2}$ and $v_{\mu}v^{\mu} \rightarrow v_{\mu}v^{\mu}/2c_{s}^{2}$, and then by working in the limit of critical temperature of the GP theory, which implies that $m^{2}\rightarrow0$, we get $v_{\mu}v^{\mu}=-1$. Thus, the radial component of the fluid four-velocity can be chosen to be the radial flow velocity (towards the hole) induced in a photon condensate, namely,
\begin{equation}
	v_{r}=-\frac{\hbar c_{0}}{\mu r}=-\xi c_{s}\frac{c_{0}}{r},
	\label{eq:velocity_SBHCL_app}
\end{equation}
where $\xi(=\hbar/\mu c_{s})$ is the correlation length, and $c_{0}(>0)$ is a constant related to the acoustic horizon's radius. Then, in order to fulfill the relation $v_{\mu}v^{\mu}=-1$, the temporal component of the fluid four-velocity should be
\begin{equation}
	v_{t}=\sqrt{1+\xi^{2} c_{s}^{2}\frac{c_{0}^{2}}{r^{2}}}.
	\label{eq:temporal_fluid_component}
\end{equation}
Thus, by introducing the dimensionless variables $r \rightarrow \xi r$ and $t \rightarrow \xi t/c_{s}$, and by assuming that the fluid is a static spherically-symmetric conservation flow, with a polar angle $\vartheta=\pi/2$, which implies that $d\vartheta=0$, we can rewrite the line element, given in Eq. (\ref{eq:acoustic_metric_app}), as
\begin{equation}
	ds^{2}=\sqrt{3}c_{s}^{2}[-f(r) dt^{2}+f(r)^{-1} dr^{2}+r^{2} d\phi^{2}],
	\label{eq:LTABH_metric_app}
\end{equation}
where the acoustic metric function $f(r)$ is given by
\begin{equation}
	f(r)=1-\frac{c_{0}^{2}}{r^{2}}.
	\label{eq:f(r)_LTABH}
\end{equation}
For simplicity and without loss of generality, we can fix $c_{s}^{2}=1/\sqrt{3}$, and hence that is the same effective metric given by Eq.~(\ref{eq:metric_SBHCL}), where $c_{0}$ is a constant related to the acoustic horizon’s radius, which can be understood as a parameter in the lab framework.

\bibliography{References}

\end{document}